\documentclass[twocolumn,showpacs,preprintnumbers,amsmath,amssymb]{revtex4}

\usepackage{graphicx}
\usepackage{dcolumn}
\usepackage{bm}

\begin{document}


\title{Induced Crystallization of Polyelectrolyte-Surfactant Complexes at the Gas-Water Interface}

\author{D. Vaknin$^1$, S. Dahlke$^1$, A. Travesset$^1$,  G. Nizri$^{2}$, and S. Magdassi$^2$}
\affiliation{$^1$Ames Laboratory and Department of Physics and Astronomy Iowa State University, Ames, Iowa 50011\\
$^2$Institute of Applied Chemistry, Hebrew University of Jerusalem, Jerusalem, Israel 91904}
\date{\today}

\begin{abstract}
Synchrotron-X-ray and surface tension studies of a strong polyelectrolyte (PE) in the semi-dilute regime ($\sim$0.1M monomer-charges) with varying surfactant concentrations show that minute surfactant concentrations induce the formation of a PE-surfactant complex at the gas/solution interface.  X-ray reflectivity and grazing angle X-ray diffraction (GIXD) provide detailed information of the top most layer, where it is found that the surfactant forms a two-dimensional liquid-like monolayer, with a noticeable disruption of the structure of water at the interface.  With the addition of salt (NaCl) columnar-crystals with distorted-hexagonal symmetry are formed.  
\end{abstract}

 \maketitle

There has been a growing interest in the phase-behavior, aggregation and precipitation of polymer-surfactant mixtures, in particular of ionic surfactants and oppositely charged flexible polyelectrolytes (PEs)\cite{Chu95,Yeh96,Hanson98,Kogej01,Hanson01,Strey03}, or semi-flexible PEs such as DNA\cite{Koltover98} or actin\cite{Wong2000}.  In addition to the fundamental interest in the principles governing phase-behavior, aggregation and precipitation of polymer-surfactant mixtures, understanding the behavior of these complex systems is crucial for technological applications concerning  detergents, paints, cosmetics, and in DNA transfection (i.e., the incorporation of exogenous DNA into a  cell)\cite{Rosen78,Goddard93,Hiemenz97} and others.  

Some important aspects regarding the formation of flexible polyelectrolyte-surfactant complexes present exciting challenges, both experimentally and theoretically.  The role played by the interface in the growth and nucleation of these complexes, for example, is to a large extent unknown, although neutron and X-ray reflectivity studies provided invaluable insight into the density profile across the interface of the PE/surfactant solutions\cite{Lu95,Stubenrauch00,Staples02,Yim2000}.  The role salt concentration has on aggregation and precipitation of PE- and PE-surfactant solutions is to a large extent an open problem\cite{Strey03}.  There are recent suggestion that at high salt concentrations, macromolecules may be over-screened by counterions, effectively reversing their charge\cite{Joanny99,Nguyen00,Dobrynin01}, and that same-charge macromolecules may attract each other in several density-regimes\cite{Gelbart00}.  

\begin{figure}[htl]                                                                                      
\includegraphics[width=2.4 in]{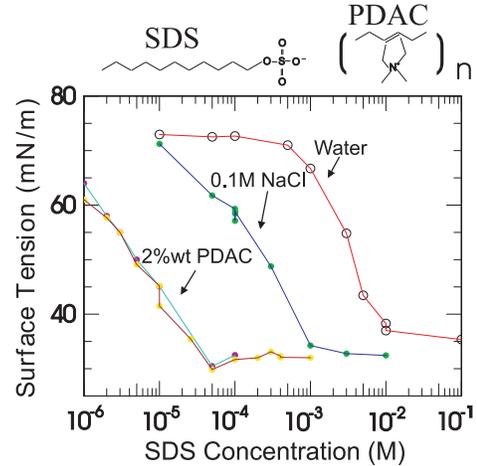}
\caption{\label{SDS_PI} Chemical structure of Sodium dodecyl sulfate (SDS) and Poly-diallyldimethylammonium chloride, (PDAC), used in this study. Surface tension as a function of SDS concentration in water and, 0.1M NaCl and 2\% PDAC both at 0.1M salt and without salt.}
\end{figure}

Previous studies on surfactant-polyelectrolyte complexes have focused on bulk properties and the self-assembly as driven by surfactant concentration. The present study focuses on interfacial behavior, in particular on the role played by the gas-water interface in the precipitation process, and how the self-assembly may be controlled by the weakening of the electrostatic interactions (salt concentration). Herein, we report surface sensitive synchrotron X-ray diffraction studies, both reflectivity and diffraction at grazing angle of incidence (GIXD), on a model system consisting of sodium dodecyl sulfate (SDS) and Poly-diallyldimethylammonium chloride (PDAC) (molecules are shown in Fig.\ \ref{SDS_PI}).   
Sodium dodecyl sulfate (SDS) (C12H27O4SNa; MW = 290 288.4) and Poly-diallyldimethylammonium chloride, (PDAC) (MW = 100,000-200,000; see Fig. 1) [C8H16NCl]n; n=685-1370;  obtained from Sigma (Materials have catalog numbers L-4509 and 40,901-4 in the 2002 Sigma-Aldrich catalog; shown in Fig.~\ref{SDS_PI}).  PDAC at 2{\%} (by weight; concentration of PE charges was 0.137M) in pure water (Milli-Q apparatus Millipore Corp., or Bedford, MA; resistivity, 18.2 M$\Omega$cm) and in 0.1M NaCl was used in all experiments.   After stirring for 10-20 minutes, solutions were poured into a temperature-controlled Teflon Langmuir trough which was maintained at 19$^o$C and enclosed in a gas tight aluminum container, where surface-tension was measured with a microbalance using a Wilhelmy filter-paper-plate.  The surface tension of the various solutions was measured as a function of time after pouring into the trough, or after compression or decompression of the surface with the Langmuir barrier, and the equilibrium value was averaged after a 30-60 min relaxation (in all cases, the time dependent pressure showed an exponential behavior, $\exp^{-t/\tau}; \tau \approx 2-5$ min.).

Surface sensitive X-ray diffraction studies of the structure of free gas/solution interface were conducted on the Ames Laboratory Liquid Surface Diffractometer at the Advanced Photon Source (APS), beam-line 6ID-B (described elsewhere\cite{Vaknin2001b}.  The highly monochromatic beam (8 keV and 16.2 keV;   $\lambda$ = 0.765334 {\AA} and $\lambda$ = 1.5498 {\AA}), selected by a downstream Si double crystal monochromator, is deflected onto the liquid surface to a desired angle of incidence with respect to the liquid surface by a second monochromator [Ge(111)and Ge(220) crystals at 8.0 and 16.2 keV respectively] located on the diffractometer.  Specular X-ray reflectivity experiments yield the electron density (ED) profile across the interface, and can be related to molecular arrangements in the film.  The ED profile across the interface is extracted by refining a slab-model that best fit the measured reflectivity by non-linear least-squares method.  The reflectivity from the slab model at a momentum transfer $Q_z$, is calculated by $ R(Q_{z})=R_{0}(Q_{z})e^{-(Q_{z}\sigma)^{2}} $
where, $R_{0}(Q_{z})$ is the reflectivity from step-like functions calculated by the recursive dynamical method, and $\sigma$ is an effective surface roughness, accounting for the smearing of all interfaces due to thermal capillary waves and surface inhomogeneities\cite{Als-Nielsen89,Vaknin2001b}.  The GIXD measurements are conducted with an evanescent incident beam that grazes the surface an angle slightly below the critical-angle for total-reflection from the surface, yielding the in-plane ordering within the penetration depth of the X-ray beam.  
\begin{figure}[ht]
\includegraphics[width=3.2 in]{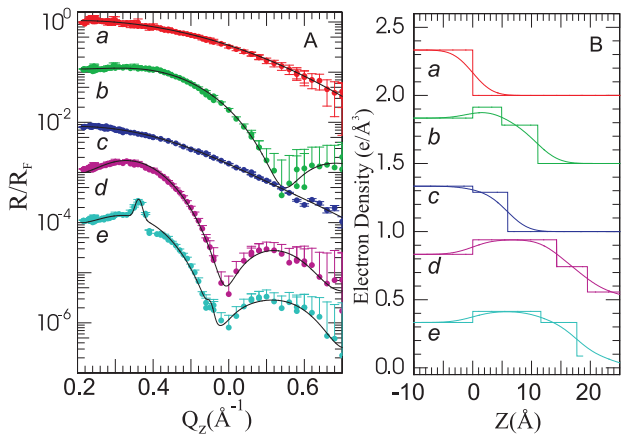}
\caption{\label{Ref1} A) Reflectivity normalized to $R_F$ ($R_F$ is the reflectivity from ideally flat water interface) for ({\it a}) 10$^{-4}M$ SDS in water ({\it b}) 10$^{-4}M$ SDS 0.1M NaCl  ({\it c})2\%wt PDAC in pure water ({\it d}) $10^{-4}M SDS$ in 2\%wt PDAC solution ({\it e}) $10^{-4}M SDS$ 2\%wt PDAC after the addition of 0.1M NaCl.   B) Electron density profiles used to generate the fitted reflectivity (solid lines in A), the step-like functions (box-model) are generated assuming no surface roughness ($\sigma=0$).}
\end{figure}

Surface tension versus SDS concentration in salt solution (0.1M NaCl) and in pure water for comparison are shown in Fig.\ \ref{SDS_PI}, and similarly for 2{\%} wt PDAC solutions in pure water and in 0.1M NaCl solution.   The reduction in surface tension indicates the adsorption of surfactants at the surface, as inferred from the Gibbs absorption equation\cite{Hiemenz97}. The onset for the reduction in surface tension at 0.1M NaCl occurs at SDS concentrations that are two orders of magnitude lower than those of SDS in pure water\cite{Tanford80}.   Likewise, in the presence of the polyelectrolyte, Figure\ \ref{SDS_PI} shows that the lowering of surface tension occurs at even lower SDS concentrations, suggestive of the formation of highly hydrophobic surfactant-polyelectrolyte complexes\cite{Ferber03,Goddard93}.   Within experimental error, our measurements in Fig.\ \ref{SDS_PI} show that surface-tension as function of SDS concentration with 2{\%} wt PDAC is not affected by the addition of simple salt (NaCl 0.1M) to water.  The lowering of surface tension by SDS to about 33 mN/m suggests that the main constituents of the top-most layer are similar with and without NaCl.

X-ray reflectivity and GIXD studies of PDAC-solutions surfaces were conducted at various SDS concentrations (with and without 0.1 M NaCl).  Figure\ \ref{Ref1} shows a sequence of normalized reflectivities, R/R$_F$ (where R$_F$ is the calculated reflectivity of an ideally flat water interface) for a typical SDS concentration, (10$^{-4}$M) in pure water and in PDAC solutions, and after adding 0.1M NaCl to the same solutions.   At SDS concentrations greater than $10^{-4}$M (in pure water), the reflectivity [Fig.\ \ref{Ref1}(A)] is similar to that of a pure water surface, although with a surface roughness $\sigma$ = 3.5 {\AA}, significantly larger than that measured for a water surface under similar conditions and with the same instrumental setup ($\sigma_W$ = 2.4 {\AA}).  The enhanced surface roughness is evidence for the presence of a dilute inhomogeneous SDS-film in a gas phase at the air/water interface.   The addition of NaCl to the SDS solution  modifies the reflectivity, and gives rise to a minimum due to the formation of a more homogeneous film, at $Q_z \approx 0.53$ {\AA}$^{-1}$.   The detailed analysis in terms of a two-box model\cite{Als-Nielsen89}, yields the ED profile shown in Fig.\ \ref{Ref1} (B) with a total film-thickness d$_{total}$ = 12 {\AA} compared to the estimated stretched SDS molecule d$_{st}$ = 19.3 {\AA}.  This implies an average molecular tilt-angle, $t$ = 51.5 degrees with respect to the surface normal [$t=\rm{acos}(d_{total}/d_{st}) $].  Additional, reflectivity studies as a function of SDS concentrations (with and without NaCl) show systematic increase of total film-thickness, up to $\approx$ 18 {\AA}, with the increase of SDS concentration\cite{VakninUnpub}.  Assuming an average SDS molecular-area $A$, the number of electrons per SDS molecule (including adsorbed molecules - H$_2$O or ions) $N_{ref}$ is given by, 
\begin{equation}
N_{ref} = N_{SDS}+N_{other} = A\int\rho(z)\rm{dz},
\label{NumEl}
\end{equation}
where  $N_{SDS}$= 148 electrons and  $N_{other}$ is the number of electrons due to integrated water molecules or ions that are inseparable from the top most layer.   Using Eq.\ (\ref{NumEl}), and the ED profiles, the lower limit for the SDS molecular area (i.e., $N_{other}$ = 0), to be $A_{min} \approx$ 35.6{\AA}$^2$, whereas assuming two bound water molecules per SDS molecule yields $A \approx$40.4{\AA}$^2$ ,  in agreement with values extracted from surface tension isotherms\cite{Prosser01}.  The two-box model shown in Fig.\ \ref{Ref1}(B) reflects the molecular ED, with a electron rich sulfate head-group compared to the low density hydrocarbon chains.  The ED of the slab associated with the hydrocarbon chains (0.27 e/{\AA}$^3$) is much lower than that of closely-packed hydrocarbon chains $\approx$ 0.34 {\AA}$^3$ in alkanes, for instance, indicating that the SDS molecules are not closely packed, most likely in a 2D liquid-like state.  The reflectivity from a 2\% PDAC in water (with no SDS) is shown in Fig.\ \ref{Ref1} to demonstrate that the effect of the PE on the surface even at this high concentration is negligible showing a small step of lower density at the interface. 
The addition of SDS to the 2\% PDAC solution brings the minimum in R/R$_F$ to $Q_z \approx 0.38$ {\AA}$^{-1}$, demonstrating the film is thicker (d$_{total}$ = 22.46 {\AA}) and more organized than that of SDS in water or in salt solution.   The thickness of the slab associated with the head-group region (d$_{head} \approx $ 11 {\AA}) is evidence that the film consists of PDAC-SDS complex at the interface.  The thickness and the ED of the two slabs at the air interface, associated with the hydrocarbon tails that are not loosely packed as in a liquid state. The most dramatic effect in the reflectivity is observed with the addition of salt to the PDAC-SDS solution, where Bragg reflections are superimposed on the reflectivity at $Q^I_z \approx 0.165$ {\AA}$^{-1}$ and at $Q^{II}_z \approx 0.345$ {\AA}$^{-1}$.  These two peaks, as will be shown below are the first and second order Bragg reflections from hexagonal structure that is normal to the scattering plane (in reflectivity configuration).  Similar neutron reflectivity studies of the dilute PDAC solutions with SDS and NaCl are consistent with the present findings\cite{Staples02}.  
\begin{figure}[]
\includegraphics[width=3. in]{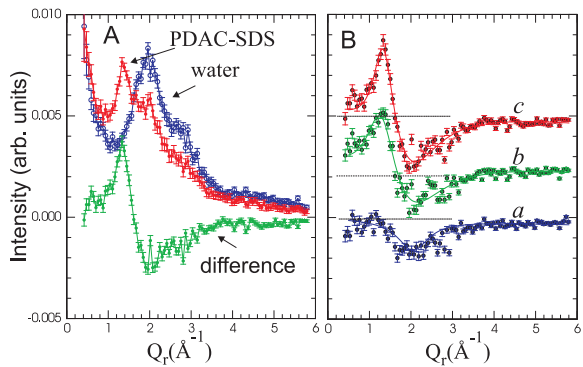}
\caption{\label{GIXD} A) In-plane diffraction (GIXD) scans of pure water, and 2\%wt PDAC 2x10$^{-5}$M SDS and 0.1M NaCl as indicated and the difference between the two scans. B) Similar differences between solutions of 2\%wt PDAC {\it a)} 0M SDS {\it b)} 2x10$^{-5}$M SDS {\it c)} 2x10$^{-4}$M SDS  . The disruption of the structure of water at the interface is apparent.} 
\end{figure}
\begin{figure}[]
\includegraphics[width=2.5 in]{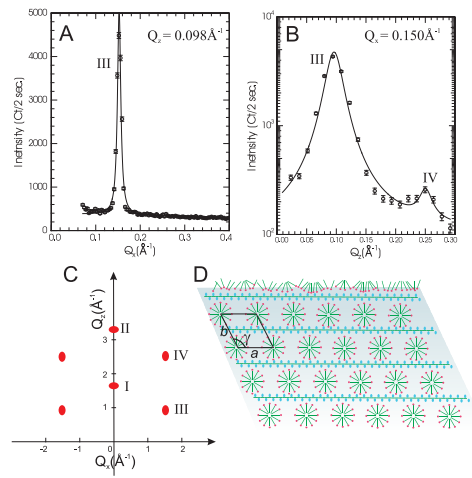}
\caption{\label{GIXD2} A) A scan along $Q_x$ of (10) peak at $Q_z$ = 0.098{\AA}$^{-1}$. B) A rod scan along the same peak (logarithmic scale) also revealing the (11) peak. C) The observed peaks in the $Q_x,Q_z$ plane D) A schematic illustration of the suggested model structure with the unit oblique cell. The long axes of cylindrical micelles are parallel to the liquid surface with each layer separated by the polymer.}
\end{figure}

The picture of a liquid-like film with disordered SDS at the gas-water interface is corroborated by our GIXD studies.  Figure\ \ref{GIXD}(A) shows the diffraction patterns of pure water surface\cite{comment1}, and that of the PDAC-SDS ($2\times 10^{-4}$M).  The difference between the two patterns is shown in Fig.\ \ref{GIXD} at two SDS concentrations.  The broad peak in the diffraction at $Q_r$ = 1.34 {\AA}$^{-1}$ of an average 4.69 {\AA} d-spacing is due to scattering from 2D-liquid hydrocarbon chains (compared to typical d-spacing for hexagonally ordered hydrocarbon chains is ~4.2 {\AA}).  The linewidth of the peak,  $\Delta Q$= 0.31 {\AA}$^{-1}$ with average correlation-length $\xi \approx 20$ {\AA}, is further evidence for the 2D disordered chains\cite{Narten90}.  

At small angles, the GIXD reveals several discrete Bragg reflections, evidence of a diffraction pattern from crystals that are highly oriented with respect to the water surface. In fact, these reflections are found to be related to those observed in the reflectivity.  As shown in Fig.\ \ref{GIXD2}, these peaks are sharp characteristic of 3D ordering, with no rod-like scattering typical of quasi-2D system.  The positions of the peaks observed and their layout as shown in Fig.\ \ref{GIXD2}(C) are consistent with a slightly distorted hexagonal structure, as depicted in Fig.\ \ref{GIXD2}(D) with {\it a} = 40.3$\pm$ 0.5 {\AA}, {\it b} = 44.6$\pm$ 0.6 {\AA}, and $\gamma $ = 121 $\pm$ 1 deg.  The peaks observed are similar although not the same as those observed by Chu and co-workers\cite{Chu95,Yeh96} in small angle X-ray scattering experiments from related systems.  Based on the anisotropy observed, we propose a simple structure of stacked cylindrical micelles, with their long axis parallel to the water-surface, where each layer is separated by a layer of the PEs, as depicted in Fig.\ \ref{GIXD2}, forming a 2D polycrystalline system with preferred orientation with respect to the surface.  
\begin{figure}[]
\includegraphics[width=2.8 in]{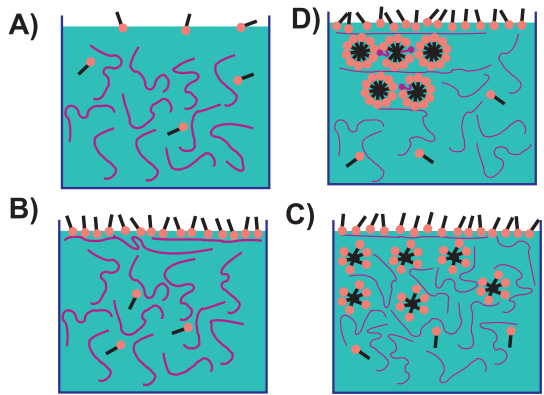}
\caption{\label{Interp} A)  Schematic pathways of complexation and subsequent crystallization A) Negligible surfactant concentration the PE is repelled from the interface. B) Below the CAC a PE surfactant is formed at the interface. C) Beyond the CAC micellization occurs D) The addition of salt transforms micelles to cylindrical shape and crystallizes them.}
\end{figure}
The distorted hexagonal diffraction pattern shows the growth is anisotropic, and implies the interface plays an important role in initiating complexation (aggregation) processes.  Another important result is the effect the addition of salt (NaCl) has on promoting PE/surfactant crystallization.    Our heuristic interpretation of the aggregation and subsequent crystallization is depicted in Fig\ \ref{Interp}.  The PE, with no surfactants or salt added, is highly soluble in water and is repelled from the air/water interface due to the discontinuity in dielectric constant\cite{Panofsky72}.  Minute increase in surfactant concentration lowers surface tension (see Fig.\ \ref{SDS_PI}) and initiates the micellization.  We argue that micelle-formation is initiated at the interface as the ideal linear bulk-separation among surfactants is $\approx$ 250 {\AA} (surfactant concentrations ($\approx $10$^{-4}$M).  The addition of NaCl to the PE/surfactant solution screens electrostatic interactions, leading to cylindrical-micelles\cite{Tanford80}.    Absorption of PE to micelles is then expected to be enhanced by the mechanism of counterion release\cite{Bruinsma98}, and we speculate that micelles will then self-attract by a similar correlation mechanism as has been recently observed in multivalent ions\cite{Angelini03}, eventually condensing into an hexagonal (columnar) crystal, and thus growing crystals from the interface.   

In summary, we have shown the fundamental role played by the interface in nucleating PE surfactant complexes, and how crystallization may be induced by salt concentration at tiny surfactant concentrations.  Our study also shows the capabilities available by X-ray scattering techniques.
 

We wish to thank M. Muthukumar and D. S. Robinson for helpful discussions. The MUCAT sector at the APS is supported by the U.S. DOE Basic Energy Sciences, Office of Science, through Ames Laboratory under contract no. W-7405-Eng-82. Measurements were supported by the U.S. DOE, Basic Energy Sciences, Office of Science, under contract no. W-31-109-Eng-38.



\begin{references}
\bibitem{comment1} For bulk water the first peak in the structure factor is observed at $Q_r$ ~ 2.0 {\AA}$^{-1}$ compared to the 1.96 {\AA}$^{-1}$ for water surface.  The penetration depth of the incident evanescent X-ray is smaller than 100 {\AA}), thereby probing the top most water layers with some contribution from bulk liquid.  The shift in the position of the first peak indicates a slightly larger O-O d-spacing at the interface.
\end{references}
\end{document}